\documentclass{article}

\usepackage{PRIMEarxiv}

\usepackage[utf8]{inputenc} % allow utf-8 input
\usepackage[T1]{fontenc}    % use 8-bit T1 fonts
\usepackage{hyperref}       % hyperlinks
\usepackage{url}            % simple URL typesetting
\usepackage{booktabs}       % professional-quality tables
\usepackage{amsfonts}       % blackboard math symbols
\usepackage{nicefrac}       % compact symbols for 1/2, etc.
\usepackage{microtype}      % microtypography
\usepackage{lipsum}
\usepackage{tabularx}
\usepackage[table]{xcolor}
\usepackage{multirow}
\usepackage{multicol}
\usepackage{fancyhdr}       % header
\usepackage{graphicx}       % graphics
\graphicspath{{media/}}     % organize your images and other figures under media/ folder

\title{Learning Class-Specific Spectral Patterns to Improve Deep Learning Based Scene-Level Fire Smoke Detection from Multi-Spectral Satellite Imagery
}

\author{
  Liang Zhao, Jixue Liu, Stefan Peters, Jiuyong Li \\
  STEM \\
  University of South Australia \\
  Adelaide, Australia\\
  \texttt{liang.zhao@mymail.unisa.edu.au, jixue.liu@unisa.edu.au} \\
  \texttt{stefan.peters@unisa.edu.au, jiuyong.li@unisa.edu.au} \\
   \And
  Norman Mueller, Simon Oliver \\
  Digital Earth Australia \\
  Geoscience Australia \\
  Canberra, Australia\\
  \texttt{norman.mueller@ga.gov.au, simon.oliver@ga.gov.au} \\
}

\begin{document}
\maketitle

\begin{abstract}
Detecting fire smoke is crucial for the timely identification of early wild fires using satellite imagery. However, fire smoke has similar spatial and spectral characteristics to other confounding aerosols, such as clouds and haze which often confuse even the most advanced deep-learning (DL) models. Nonetheless, these aerosols also present distinct spectral characteristics in some specific bands, and such spectral patterns are useful for distinguishing the aerosols more accurately. For example, early research in satellite imagery based smoke detection tried to derive various threshold values from the reflectance and brightness temperature in specific spectral bands to differentiate smoke and cloud pixels, based on their distinct spectral characteristics in these bands. However, such threshold values were determined based on domain knowledge and are hard to generalise. In addition, such threshold values were manually derived from specific combination of bands to infer spectral patterns, making them difficult to be employed in deep learning models. In this paper, we introduce a DL module called input amplification (InAmp) which is designed to enable DL models to learn class-specific spectral patterns automatically from multi-spectral satellite imagery and improve the fire smoke detection accuracy. InAmp can be conveniently integrated with different DL architectures. We evaluate the effectiveness of the InAmp module on different Convolutional neural network (CNN) architectures using two satellite imagery datasets: USTC\_SmokeRS, derived from Moderate Resolution Imaging Spectroradiometer (MODIS) with three spectral bands; and Landsat\_Smk, derived from Landsat 5/8 with six spectral bands. Our experimental results demonstrate that the InAmp module improves the fire smoke detection accuracy of the CNN models. Additionally, we visualise the spectral patterns extracted by the InAmp module using test imagery and demonstrate that the InAmp module can effectively extract class-specific spectral patterns.
\end{abstract}

% keywords can be removed

\section{Introduction}
\label{intro}
Detecting early fire smoke from satellite imagery is recognised as a more effective and timely approach to preventing fire disasters than detecting fires directly. This is due to the fact that smoke plumes are typically the first visible indicators of wildfires from the space. By detecting fire smoke, fires can be identified when they are still small and burning at lower temperatures, such as grass fires. 
  
\begin{figure}[!t]
\centering
\includegraphics[width=4in]{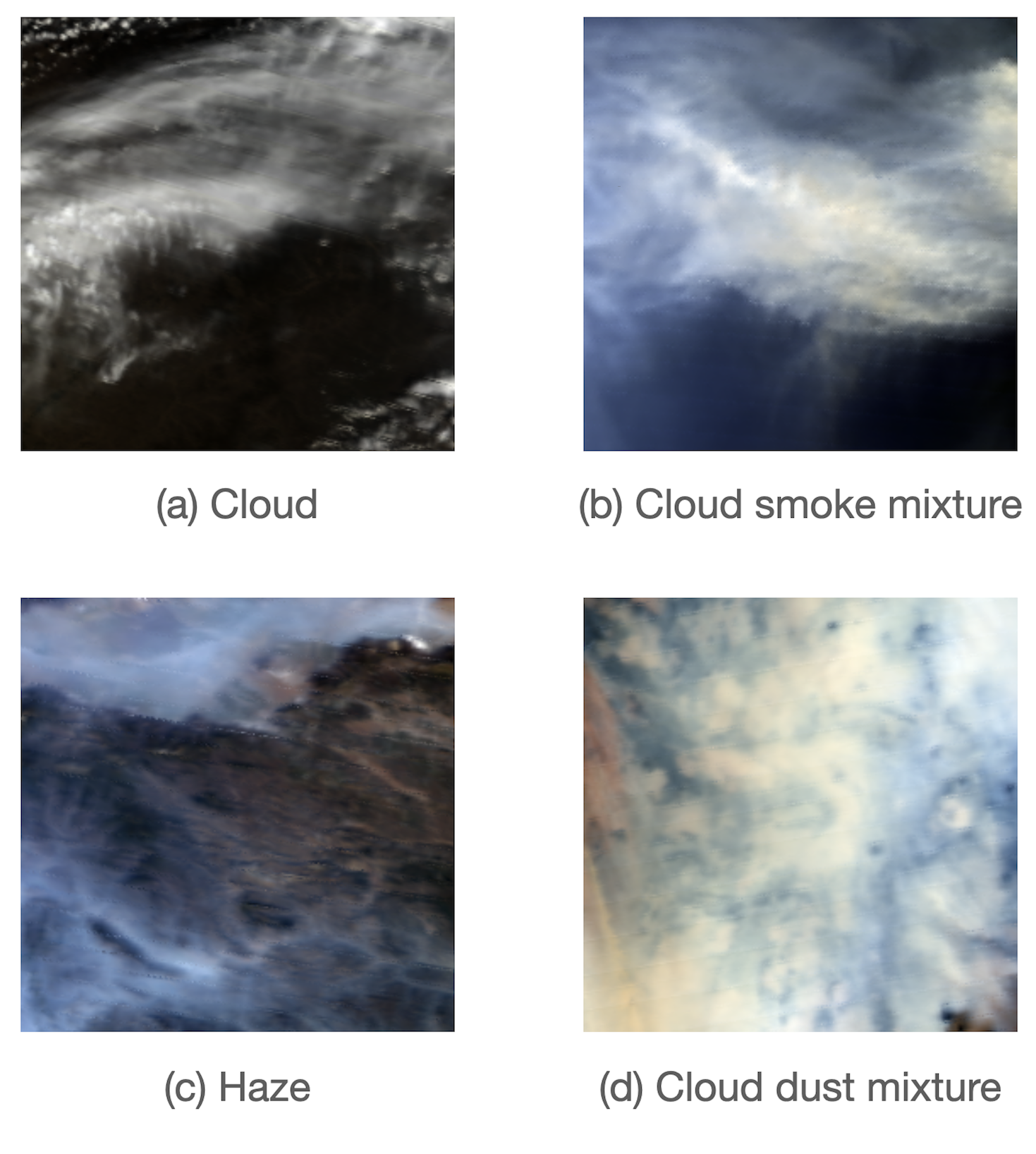}
\caption{Cloud, haze, dust, and smoke captured in Moderate Resolution Imaging Spectroradiometer (MODIS) true-colour imagery are difficult to be visually differentiated.}
\label{modis_aerosol}
\end{figure}

However, fire smoke detection presents its own set of challenges. For example, fire smoke shares similar spatial and spectral characteristics with other aerosols (e.g. cloud, fog, haze, and dust), and often intermingles with these aerosols in satellite imagery, making the accurate detection of fire smoke amongst these aerosols extremely challenging, as demonstrated in Figure~\ref{modis_aerosol}. 

Fire smoke detection methods from satellite imagery can be broadly classified into two categories: pixel-level and scene-level. In pixel-level detection, the goal is to segment all fire smoke pixels from other pixels in the image. In scene-level detection, on the other hand, the aim is to detect fire smoke at a higher level, by classifying the entire image as either ``Smoke'' or other classes based on whether the scene captured in the imagery contains fire smoke.

Early research on fire smoke detection from satellite imagery primarily focused on the pixel-level. Researchers used mathematical and statistical methods to derive threshold values from the reflectance and brightness temperature (BT) values in multiple spectral bands for each pixel. Smoke pixels were then distinguished from other pixels based on the differences in these spectral-band threshold values \cite{christopher1996first,baum1999grouped,asakuma2002detection,chrysoulakis2003new,chrysoulakis2007improved,shukla2009automatic,ismanto2019classification}, which can be interpreted as some specific spectral patterns. However, such threshold values were handcrafted based on domain knowledge and experience, and may be influenced by local conditions. This makes the threshold values difficult to be generalised.

In the last decade, deep-learning (DL) has significantly advanced both pixel-level and scene-level fire smoke detection from satellite imagery. DL models like Convolutional Neural Networks (CNNs) and Vision Transformers (ViTs) can automatically extract highly abstract features without the need of cumbersome feature engineering, and have been proven to have greatly improved the fire smoke detection accuracy.  %These operations have been successful in extracting spatial patterns hidden in neighboring pixels.

However, one notable limitation of CNNs and ViTs is that they fail to account for spectral patterns that are indicative for pixels belonging to different aerosols. CNN models focus on spatial feature extraction and do not explicitly extract pixel-level spectral features from the input layer, while ViTs focus more on correlations of sub-regions in features due employing the self-attention mechanism.
%Such spectral patterns, like the threshold values, must be calculated for each pixel so that neighbouring pixels belonging to different classes can be distinguished. Unfortunately, based on our observation, the State-of-the-art CNNs, regardless of whether they are used for pixel-level segmentation or scene-level classification, typically start with spatial feature extraction in their first convolutional layers by using $3 \times 3$ or larger filters. Such filters examine the associations of neighbouring pixels instead of focusing on spectral characteristics of individual pixels. ViTs has the same issue since they take transformed high-level features extracted by CNNs or through patch embedding as the input to the encoder, and such features do not examine pixel-level spectral patterns in the original imagery. 

\begin{figure}[!t]
\centering
\includegraphics[width=6in]{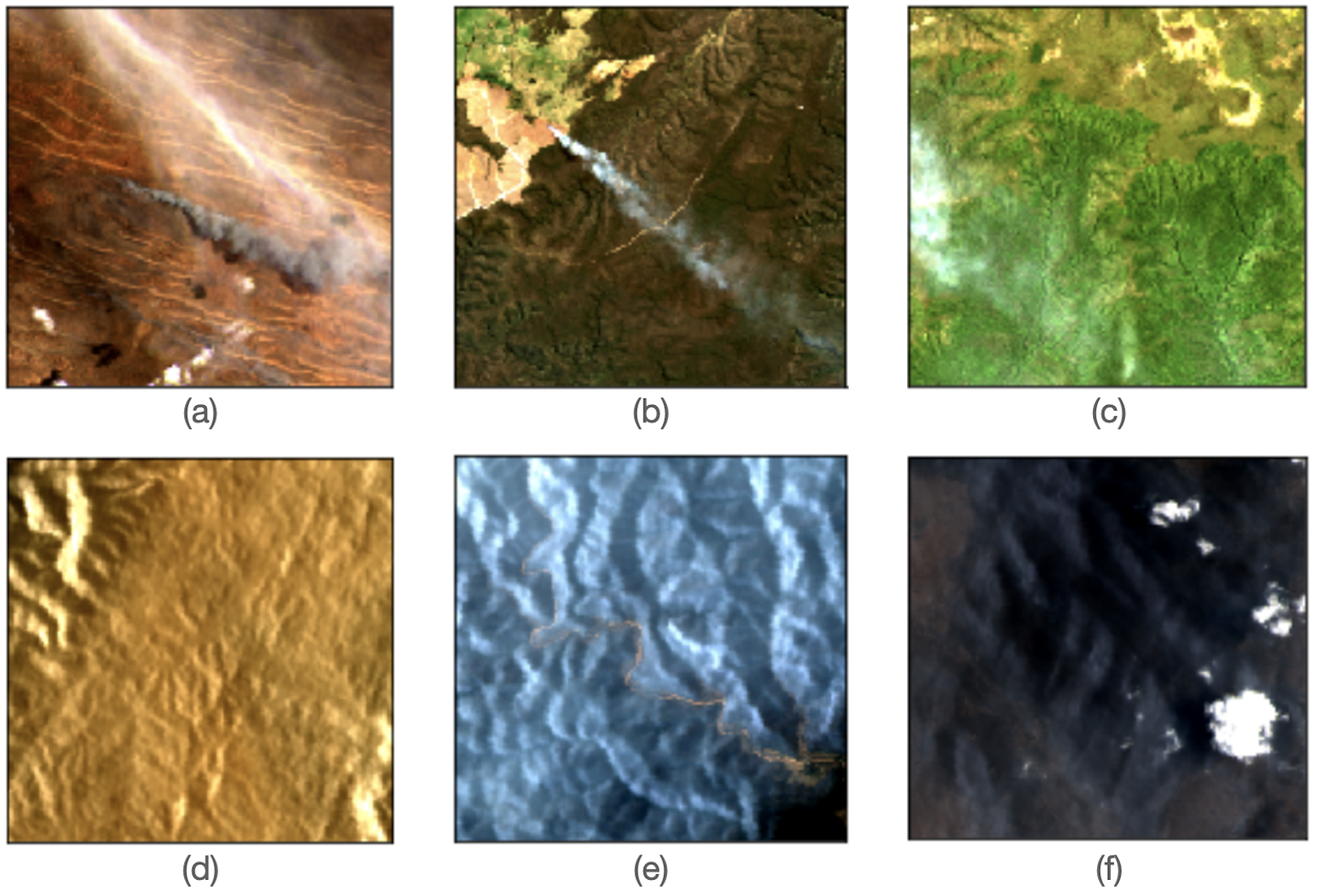}
\caption[Variants of fire smoke in Landsat 8 OLI true-colour imagery. (a) Dark grey fire smoke plumes under cirrus clouds. (b) Long slim fire smoke plume in bright colour. (c) Dispersed fire smoke on the edge of the image. (d) Brown-coloured dense fire smoke in the whole image. (e) Wide, dispersed fire smoke in light blue colour covering the whole image. (f) Spread dense fire smoke in dark grey colour under altocumulus clouds.]{Variants of fire smoke in Landsat 8 OLI true-colour imagery. (a) Dark grey fire smoke plumes under cirrus clouds. (b) Long slim fire smoke plume in bright colour. (c) Dispersed fire smoke on the edge of the image. (d) Brown-coloured dense fire smoke in the whole image. (e) Wide, dispersed fire smoke in light blue colour covering most of the image. (f)) Dense fire smoke in dark grey colour under altocumulus clouds. Adopted from ~\cite[Figure~1]{zhao2022investigating}}
\label{landsat_smk}
\end{figure}

One reasonable hypothesis is that the accuracy of the DL models for fire smoke detection can be benefited by explicitly extracting pixel-level spectral patterns at the very beginning of the model, particularly when using multi-spectral satellite imagery. As demonstrated in Figure~\ref{landsat_smk} \cite[Figure~1]{zhao2022investigating}, the shapes and colours of fire smoke can vary greatly, indicating that spatial patterns are insufficient for accurate fire smoke detection from satellite imagery.

%It is reasonable to believe that the accuracy of the fire smoke detection DL models could be improved provided useful pixel-level spectral patterns can be effectively explored, especially when using multi-spectral satellite imagery. As demonstrated in Figure~\ref{landsat_smk} \cite[Figure~1]{zhao2022investigating}, the shapes and colours of fire smoke can vary greatly, indicating that spatial patterns are insufficient for accurate fire smoke detection from satellite imagery.

To validate the hypothesis, we designed a module %To assist training DL models for accurately detecting fire smoke from satellite imagery with limited training data, 
named input amplification (InAmp) to enable a DL model to automatically learn class-specific pixel-level spectral patterns that are useful for distinguishing fire smoke from the background and other visually similar objects. Unlike the threshold values that were derived based on domain knowledge and cannot be generalised, the spectral patterns extracted by InAmp are automatically learned through supervised training without human intervention. %The InAmp bridges the gap and explores pixel-level spectral pattern for the first time in the literature of DL-based scene-level fire smoke detection using multi-spectral satellite imagery.

We compared the performance of various baseline CNNs trained with and without the InAmp module to evaluate its effectiveness. The baseline CNNs are ResNet50 \cite{7780459}, InceptionResnetV2 \cite{szegedy2017inception}, MobileNetV2 \cite{sandler2018mobilenetv2}, and VIB\_SD \cite{zhao2022investigating}. We trained the models with two satellite imagery datasets, namely USTC\_SmokeRS \cite[dataset]{ba2019smokenet} and a multi-spectral Landsat imagery dataset (referred to as Landsat\_Smk herein after) \cite[dataset]{zhao2022investigating}. The former dataset consists of RGB (i.e. the visible bands) imagery from MODIS, while the latter dataset consists of imagery from Landsat 5 and Landsat 8 with six spectral bands including the RGB bands, the Near InfraRed (NIR) band, and the Shortwave Infrared (SWIR) bands one (SWIR\_1) and two (SWIR\_2).

Our results showed that incorporating the InAmp module effectively improved the prediction accuracy of different CNN architectures for both datasets. 
 
The novelty and the contributions of the InAmp module can be interpreted as follows:

\begin{itemize}
    \item InAmp bridges the gap and enables DL models to explore pixel-level spectral patterns alongside spatial features, for the first time in the literature, for scene-level fire smoke detection using multi-spectral satellite imagery.
    \item InAmp is lightweight, and can be conveniently integrated with existing DL models (e.g., CNNs or ViTs using high-level features extracted from CNNs) with little overhead on the computational complexity;
    \item InAmp extracts pixel-level spectral patterns in a task-driven manner, producing class-specific patterns that can be visualised to aid in the interpretation of DL models;
    \item InAmp has the potential to be applied in various domains beyond fire smoke detection. For instance, it can be utilised for water observation, vegetation disease detection, and other related fields; 
    \item InAmp may be used to facilitate transfer learning for cross-sensor model training.
\end{itemize}

The following content of this paper is organised as follows:
Section~\ref{related_work} provides a review of related work. Section~\ref{inamp} introduces the proposed InAmp module. Section~\ref{experiment} describes the experimental settings, including datasets, training settings, and evaluation metrics. Section~\ref{results} interprets the results, including ablation studies and the parameter selection of the InAmp module. Section~\ref{discussion} discusses potential applications of InAmp and our future work. Section~\ref{conclusion} presents the conclusion. 

\section{Related Work} 
\label{related_work}

In this section, we provide a review of approaches to fire smoke detection from satellite imagery, with a focus on pixel-level and scene-level methods in
section~\ref{subsec:pixel-approches} and section~\ref{subsec:scene-approches}, respectively. Additionally, we provide a general context in terms of spectral pattern, which is discussed in section~\ref{subsubsec:spec_pat}, based on the literature.

\subsection{Pixel-level Fire Smoke Detection} \label{subsec:pixel-approches}
\subsubsection{Spectral Pattern} \label{subsubsec:spec_pat}
A ``Spectral pattern'' can be generally described as a pattern in the pixel values across the spectral bands of remotely sensed imagery, and is typically related to the spectral signature of a certain surface feature. Although ``Spectral pattern'' has been frequently mentioned in the areas of remote sensing and computer vision \cite {lavine2002genetic,fingelkurts2003systematic,duong2012water,bazar2016nir}, to the best of our knowledge, its definition has been vague in the literature. 

Spectral indices, such as Normalised Difference Vegetation Index (NDVI), Normalised Burn Ratio (NBR), and Normalised Difference Built-Up Index (NDBI), can be considered a special type of spectral pattern. Spectral indices are calculated using designated spectral bands and formulas, and their values present different patterns against the original bands, indicating the existence or occurrence of certain objects or events.

Table~\ref{spectral_indices} displays some of the common spectral indices used in remote sensing literature. It should be noted that spectral patterns should be indicative to the type of a pixel and are usually not simple linear combinations of the selected band values. We will give a formal definition of spectral pattern in section~\ref{inamp} to better reflect its implications in this paper.

\begin{table*}
\renewcommand{\arraystretch}{1.8}
\caption{Some spectral indices used in remote sensing}
\label{spectral_indices}
\centering
\begin{tabular}{c c c c}
\toprule
\textbf{Index} & \textbf{Formula} & \textbf{Objective} & \textbf{References}\\
\cmidrule{1-4}
\multirow{4}{*}{NDVI} & \multirow{4}{*}{$\frac{NIR-Red}{NIR+Red}$} & \multirow{4}{*}{Highlight vegetation} & \cite{pettorelli2005using}\\
& & & \cite{fensholt2009evaluation}\\
& & & \cite{escuin2008fire}\\
& & & \cite{huang2021commentary} \\
\cmidrule{1-4}
\multirow{3}{*}{NBR} & \multirow{3}{*}{$\frac{NIR-SWIR}{NIR+SWIR}$} & \multirow{3}{*}{Highlight burnt areas} & \cite{escuin2008fire}\\
& & & \cite{berndt2019towards}\\
& & & \cite{que2019analisis} \\
\cmidrule{1-4}
\multirow{3}{*}{NDBI} & \multirow{3}{*}{$\frac{SWIR-NIR}{SWIR+NIR}$} & \multirow{3}{*}{Highlight urban areas} & \cite{ali2019monitoring}\\
& & & \cite{prasomsup2020extraction}\\
& & & \cite{zheng2021improved}\\
\bottomrule
\end{tabular}
\end{table*}

\subsubsection{Approaches to Pixel-level Fire Smoke Detection} 

Early approaches to fire smoke detection from satellite imagery primarily relied on statistical or traditional machine learning methods. These methods aimed to identify fire smoke pixels by setting multiple spectral-band threshold values based on the fact that the spectral signatures of fire smoke often exhibit distinctive patterns from clouds and other confounding objects in certain spectral bands. The threshold values were calculated using designated spectral bands and formulas, much like calculating spectral indices. For example, \cite{christopher1996first} used multiple spectral-band threshold values to extract texture features related to levels of gray color for marking smoke pixels. \cite{baum1999grouped} proposed a grouped threshold approach to discern smoke, snow, cloud, fire, and clear sky. \cite{chrysoulakis2003new} and \cite{chrysoulakis2007improved} further used experimental thresholds to measure the multi-temporal and multi-spectral change in four derived pseudo-bands for fire smoke detection. \cite{asakuma2002detection} proposed a supervised Euclidean classification model for smoke detection based on extracted texture features using multiple spectral-band threshold values. \cite{ismanto2019classification} employed a supervised classification tree model to classify pixels in Himawari-8 imagery into "Smoke" and six other background or confounding classes using multiple spectral-band threshold values. However, a significant drawback of these methods is that the spectral-band threshold values were manually derived based on experience and domain knowledge, and can vary greatly for different sensors or environmental conditions.

Later, neural networks were also explored for pixel-level fire smoke detection. \cite{li2001automatic} proposed a simple neural network with one hidden layer and 10 neurons to classify pixels in Advanced Very-High-Resolution Radiometer (AVHRR) imagery. The model took pixel-level reflectance values or BT values from five spectral bands as input and classified the pixels into three classes: "Smoke", "Cloud", and "Land". \cite{li2015forest} proposed another shallow neural network consisting of one hidden layer and 20 neurons to classify pixels in MODIS imagery into "Smoke", "Cloud", and "Underlying Surface". The model's input vector consists of six features containing various reflectance and BT information derived from different spectral bands. Both models could automatically extract features from the input spectral/pseudo bands and achieved good accuracy, and they demonstrated the importance of spectral patterns in fire smoke detection. However, the input features were determined based on domain knowledge, and spatial information could not be explored from individual pixels.

To incorporate spatial information for pixel-level fire smoke detection, \cite{larsen2021deep} proposed a more advanced DL model: a fully convolutional neural network (FCN) for smoke segmentation in Himawari-8 imagery. It is a noteworthy example of CNN-based methods utilising manipulated spectral patterns for fire smoke detection. The imagery data used to train the model consists of seven channels, including six spectral bands and a predefined spectral index: the fire radioactive power (FRP). While FRP is considered a reasonable indicator for fire smoke detection, the authors did not explore its contribution to the model's accuracy. Additionally, the model does not extract useful spectral patterns but, instead, arbitrarily uses one spectral pattern among many other possible choices. 

\subsection{Scene-level Fire Smoke Detection} \label{subsec:scene-approches}

Scene-level fire smoke detection from satellite imagery has been based on DL models, predominantly CNNs. The convolutional layers in CNNs typically learn $3 \times 3$ or larger filters to weight the neighbouring pixels based on their spatial importance. Consequently, scene-level fire smoke detection from satellite imagery using CNNs tend to focus more on spatial patterns.

Specifically designed CNN models for fire smoke detection have been shown to outperform other CNN models due to the unique challenges associated with the task \cite{ba2019smokenet,chen2021global2salient}. SmokeNet \cite{ba2019smokenet}, the first CNN model designed for scene-level fire smoke detection from satellite imagery, was trained on the USTC\_SmokeRS dataset which was proposed in the same work. SAFA \cite{chen2021global2salient}, the current SOAT CNN model, was also trained using the USTC\_SmokeRS dataset.
\cite{zhao2022investigating} recently proposed a lightweight CNN model VIB\_SD which achieved comparable accuracy to SAFA while using less than 2\% of its number of parameters. VIB\_SD was trained on the six-band Landsat\_smk dataset, which was also constructed by the authors from Landsat multi-spectral imagery, to investigate using additional IR bands for early fire smoke detection.

Like other CNN architectures, SmokeNet, SAFA, and VIB\_SD all begin with spatial feature extraction using $3 \times 3$ or larger filters, and this means that the models do not directly examine pixel-level spectral patterns.

It is noteworthy that all three models employed the attention mechanism which is also a key component of the proposed InAmp module. The attention mechanism is inspired by the human cognitive system which gives higher priority to distinctive components when processing information from multiple sources \cite{tsotsos1995modeling}. Readers can refer to \cite{niu2021review} for a more detailed explanation of the attention mechanism.

The proposed InAmp module applies the attention mechanism from a novel perspective. Unlike previous applications of attention mechanisms that only focus on spatial features, channels within a feature map, or spectral bands, the InAmp module is intentionally designed to identify associations amongst spectral bands at the pixel level, in conjunction with spatial information. The InAmp module aims to automatically extract class-specific pixel-level spectral patterns and integrate them with spatial patterns, ultimately enhancing DL-based approaches for detecting fire smoke at the scene-level from satellite imagery.

The InAmp module functions as an input pre-processing block within a DL model and therefore can be also easily incorporated into ViTs. The module's architecture will be outlined in the following section.

\section{InAmp} \label{inamp}

``Spectral pattern'' has been used in the previous work, but there is not a uniform definition. To make discussions precise in this paper, we first give a formal definition of ``spectral pattern'' as follows.

Given an input image $X \in \mathbb{R}^{W \times H \times C}$, where $W \in \mathbb{N}$, $H \in \mathbb{N}$, and $C \in \mathbb{N}$ represent the width, height, and number of spectral channels of $X$, a spectral pattern $f$ of $X$ refers to a semantic mapping that transforms the original values in each spectral channel of any pixel $P_{i,j} \in \mathbb{R}^C$ in $X$ to one new value $P^f_{i,j} \in \mathbb{R}$:
\begin{equation}
\label{eq:spectral_pattern_def}
    f \colon  (P^1_{i,j}, \ldots, P^k_{i,j}, \ldots, P^C_{i,j}) \longmapsto P^f_{i,j}
\end{equation}
where $i \in [0,W)$, $j \in [0, H)$ are the indices of the pixel $P_{i,j}$; $P^k_{i,j} \in \mathbb{R}$ is the value of the pixel $P_{i,j}$ in the $k$th channel, and $k \in [1,C]$.

\begin{figure*}[!h]
\centering
\includegraphics[width=6in]{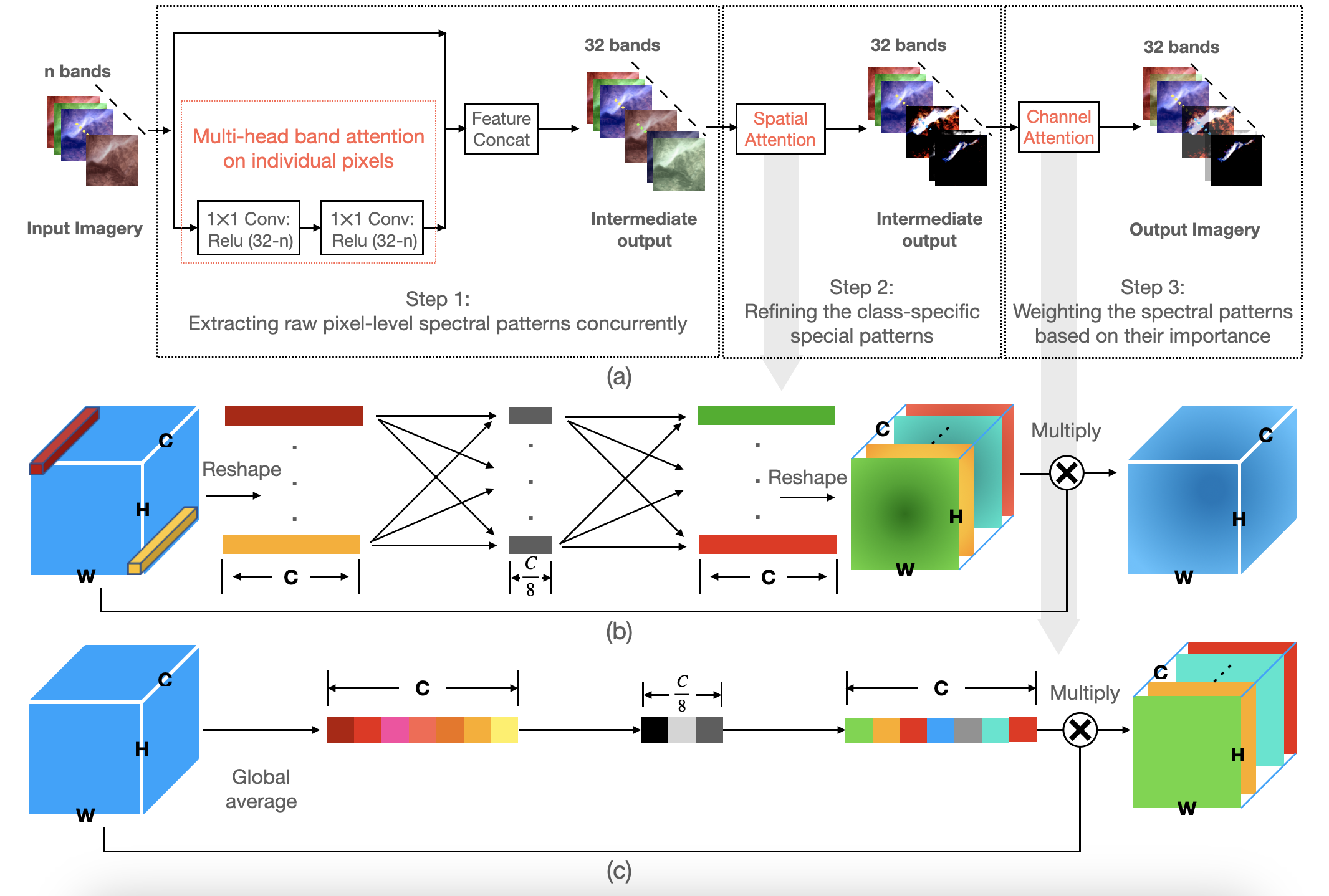}
\caption[The Structure of the InAmp Module. (a) InAmp (b) Spatial Attention (c) Channel Attention.]{The architecture of the InAmp module. (a) InAmp. (b) Spatial Attention. 
(c) Channel Attention.}
\label{InAmp_structure}
\end{figure*}

We now elucidate how the InAmp module extract and integrate pixel-level spectral patterns with spatial patterns to facilitate scene-level fire smoke detection from satellite imagery using DL models. Figure~\ref{InAmp_structure} (a) depicts the structure of the InAmp module which has been devised to serve the following objectives:
\begin{enumerate}
    \item Automatic extraction of pixel-level spectral patterns;
    \item Extraction of multiple spectral patterns concurrently, with a focus on those containing valuable class-specific information;
    \item Integration of spectral patterns and spatial patterns to enhance the accuracy of fire smoke detection.
\end{enumerate}

The InAmp module accomplishes the above objectives by means of three successive steps that employ different types of attention:
\begin{enumerate} 
    \item Band attention applied to individual pixels across the input spectral bands, which produces raw spectral patterns.
    \item Spatial attention employed on pixels within each spectral band and spectral pattern, leading to the refinement of spectral patterns.
    \item Channel attention utilised on the combined spectral bands and spectral patterns, focusing on significant spectral patterns.
\end{enumerate}

In the first step, the InAmp module employs $1 \times 1$ filters with the Relu activation function to calculate the attention weights of a pixel's spectral bands. The $1 \times 1$ filter $F=[F^1, \dots, F^k, \dots, F^C](F^k\in \mathbb{R})$ linearly maps any pixel $P_{i,j}=[P_{i,j}^1, \dots, P_{i,j}^k, \dots, P_{i,j}^C](P_{i,j}^k\in \mathbb{R})$ to a new value $P_{i,j}^F=\sum_{k=1}^C F^k P_{i,j}^k$. The Relu function then applies a non-linear transformation and produces a new value $P_{i,j}^{Relu(F)}$. The above process aims at extracting spectral patterns as in equation~\ref{eq:spectral_pattern_def}. We use multiple $1 \times 1$ filters to achieve multi-head band attention on the pixels. When $N$  filters are used in a $1 \times 1$ Conv2D layer with the Relu activation function, $N$ spectral patterns can be extracted concurrently. Two such $1 \times 1$ Conv2D layers are stacked to extract the spectral patterns. The output feature maps of the two Conv2D layers represent the raw spectral patterns. They all have the same dimension as the spectral bands in the input imagery, and therefore can be treated as deep-pseudo bands and concatenated with the original spectral bands. By doing so, the spatial and spectral information in the original imagery can be preserved in the following steps for the refinement of the extracted spectral patterns. During the backpropagation learning process, the weights of the pixels in the spectral patterns are adjusted based on their contributions to the classes of images. For convenience, the extracted spectral patterns are called deep-pseudo bands. 

Since CNNs typically require a fixed number of channels in the input, for the sake of convenience in implementation, we set the number of output channels of the InAmp module instead of setting the number of $1 \times 1$ filters used. In our case, the number of output channels is set to $32$, so the number of $1 \times 1$ filters used in the two Conv2D layers is $32-n$,  where $n$ is the number of spectral bands in the input imagery. We chose the number $32$ based on the results of ablation studies while considering the computational complexity.

In the second step, the InAmp module incorporates spatial attention to refine the extracted spectral patterns. This attention mechanism learns the importance of each pixel in each spectral or deep-pseudo band based on its spatial location, sharpens the distribution of pixels belonging to the same target class and makes the spectral patterns distinctive for each target class. For instance, if a spectral pattern is associated with smoke, the spatial attention will enhance the smoke pixels while suppressing pixels belonging to other classes in the same spectral pattern, or vice versa.

In the third step, the InAmp module proceeds to apply channel attention to weight the importance of the extracted spectral patterns before they are used as input to a DL model. This attention mechanism assigns higher weights to the spectral patterns that have more representative power to the target classes, and guides the model to learn class-specific patterns effectively. As a result, the DL model can accurately distinguish between different classes of targets. 

The implementation of spatial attention and channel attention is normal and typified by \cite{ba2019smokenet} and \cite{zhao2022investigating} in smoke detection. In this paper, the spatial attention and channel attention are applied in a novel perspective so that they, jointly with $1 \times 1$ filters, produce refined pixel-level spectral patterns while preserving spatial features and emphasising class-specific information. Whereas, previous applications of spatial attention and channel attention served for enhancing the extraction of spatial patterns. 

The InAmp module learns class-specific spectral patterns from various band combinations automatically, without requiring human expertise or intervention. This is in contrast to traditional methods, such as spectral indices and threshold values which rely on predefined formulas based on experience and domain knowledge. Additionally, the InAmp module automates the link between the extracted spectral patterns and the scene-level classification, streamlining the entire process and making it more efficient.

\section{Experimental settings} \label{experiment}

In this section, we will introduce the datasets used in this paper, the training settings, the evaluation metrics.

\subsection{Datasets}

To the best of the authors' knowledge, there are only two scene-level labeled fire smoke satellite imagery training datasets available: the aforementioned USTC\_SmokeRS dataset and the Landsat\_smk dataset. The USTC\_SmokeRS dataset consists of RGB bands, while the Landsat\_smk dataset contains six bands, including RGB, NIR, SWIR\_1, and SWIR\_2. Both datasets are utilised in this paper to explore the effectiveness of the InAmp module.

Table~\ref{img_datasets} shows the basic information about the two training datasets.

\begin{table}%[width=.9\linewidth,cols=4,pos=h]
\renewcommand{\arraystretch}{1.3}
\caption{A summary of datasets used in this paper}
\label{img_datasets}
\centering
\begin{tabular}{c c c c}
\toprule
\multirow{2}{*}{\textbf{Dataset}} & \multirow{2}{*}{\textbf{Classes}} & \textbf{\#Images} & \textbf{\#Images}\\
& & \textbf{/Class} & \textbf{Total}\\
\midrule
\multirow{6}{*}{USTC\_SmokeRS} 
& Cloud & 1164 & \multirow{6}{*}[-.8em]{6225}\\\cmidrule{2-3}
& Dust & 1009 \\\cmidrule{2-3}
& Haze & 1002 \\\cmidrule{2-3}
& Land & 1027 \\\cmidrule{2-3} 
& Seaside & 1007 \\\cmidrule{2-3}
& Smoke & 1016 \\\cmidrule{1-4}
\multirow{3}{*}{Landsat\_smk} 
& Clear & 616 & \multirow{3}{*}{1836}\\\cmidrule{2-3}
& Other\_aerosol & 605 \\\cmidrule{2-3}
& Smoke & 615 \\%\cmidrule{1-4}
\bottomrule
\end{tabular}
\end{table}

\subsection{Training Settings}

We selected four baseline models , namely ResNet50 \cite{7780459}, InceptionResnetV2 \cite{szegedy2017inception}, MobileNetV2 \cite{sandler2018mobilenetv2}, and VIB\_SD \cite{zhao2022investigating}, with different depths and various numbers of parameters. VIB\_SD is a lightweight model designed particularly for fire smoke detection, other models are well-known in the literature. We aim to verify whether the InAmp module can be used for different CNN architectures with variant depths and numbers of parameters. 

The baseline models were firstly trained with the two datasets. We then inserted the InAmp module right after the input layer. The output of the InAmp module has the same width and height as the original input imagery, but has 32 channels with extracted spectral patterns and is fed to the  baseline models as the new input.  

For both datasets, we used 64\% of the data for training, 16\% for validation, and 20\% for test. The test results were used for comparison. To minimise the risk of overfitting, all input imagery in the training data were augmented with random horizontal and vertical flipping. We kept the augmentation simple to avoid introducing noise.
 
All models were trained using an input size of $256 \times 256$ which is the original size of the image files. It is important to note that ResNet50 and InceptionResNetV2 have default input sizes of $224 \times 224$ and $299 \times 299$, respectively. Using these default input sizes requires resizing the input imagery, which can introduce interpolated pixel values across all the input bands. This can cause the learned spectral patterns to deviate significantly from the true spectral patterns. To avoid this issue, we changed the input size of ResNet50 and InceptionResNetV2 to $256 \times 256$. This allowed the models to learn effectively without introducing interpolated pixels, and did not impact the comparison.

We set the batch size to 32, and set the number of epochs to 300. To avoid redundant training, early stopping was applied given the validation accuracy does not increase within 60 epochs. The learning rate was set to 0.01 initially and reduced by a factor of 0.8 if the validation loss does not drop within 20 epochs. We used the Adam optimiser \cite{kingma2014adam} for the optimisation.

All algorithms were implemented using TensorFlow and trained under the Ubuntu 16.4 operation system. We used a global random seed together with necessary local seeds for preparing the datasets and training the models with mirror strategy using two Nvidia Gforce 1008 GPUs. This is to make the models more comparable considering there are many random processes during the training, such as random parameter initialisation, random dataset splitting and shuffling, random job assignment, etc.

The results of each baseline model were compared to the results of the same model that incorporated the InAmp module.

\subsection{Evaluation Metrics}
We adopted accuracy (\%) and kappa-coefficient (Kappa) as evaluation metrics as used in \cite{ba2019smokenet,chen2021global2salient,zhao2022investigating}. Since False Negative (FN) is also an important metric for natural disaster detection, we further added FN of the target class ``Smoke'' as an evaluation metric.

\section{Experimental Results} \label{results}

In this section, we will present and compare the test results of all the baseline models with or without the InAmp module in section~\ref{result_comp}. We will visualise and analyse some of the spectral patterns in the deep-pseudo bands extracted by the InAmp module in section~\ref{InAmp_patterns}. Finally, we will show the results of the ablation studies in section~\ref{ablation}

\subsection{Model Performance with/without InAmp} \label{result_comp}

The test results of the baseline models with and without the InAmp module using the USTC\_SmokeRS dataset and the Landsat\_smk dataset are shown in Tabel~\ref{results_USTC} and Tabel~\ref{results_Landsat} respectively. 

The results showed that adding the InAmp module only slightly increased the parameter numbers (\#Params) compared with the original models. 

Training with the USTC\_SmokeRS dataset, the InAmp module effectively improved all three evaluation metrics for ResNet50, InceptionResnetV2, and VIB\_SD, whereas, the original MobileNetV2 had better results for all three evaluation metrics.  

Training with the Landsat\_smk dataset, all four baseline models gained significant improvement in terms of accuracy and kappa coefficient and, except ResNet50, all other three baseline models gained improvement in terms of FN rate for the class ``Smoke''.

The above results show that the InAmp module can effectively improve CNN-based fire smoke detection from satellite imagery. We suspect that the compromised performance of MobileNetV2 with InAmp when trained using the USTC\_SmokeRS dataset is related to the extensive use of $1 \times 1$ filters in its depth-wise separable convolution and the inverted residual blocks.

\begin{table*}
\renewcommand{\arraystretch}{1.1}
\setlength\tabcolsep{4pt}
\caption{Results of using the USTC\_SmokeRS dataset}
\label{results_USTC}
\centering
\begin{tabular}{c c c c c c}
\toprule
\textbf{Model} & \textbf{InAmp} & \textbf{\#Params} & \textbf{Accuracy} & \textbf{Kappa} & \textbf{FN}\\
\midrule
\multirow{2}{*}{ResNet50} & No & 23.60M & 86.43\% & 83.71\% & 21.60\%\\\cmidrule{2-6}
& Yes & 23.69M & \cellcolor{lightgray}88.67\% & \cellcolor{lightgray}86.41\% & \cellcolor{lightgray}18.31\%\\
\midrule
\multirow{2}{*}{InceptionResnetV2} & No & 54.35M & 88.27\% & 85.92\% & 16.43\%\\\cmidrule{2-6}
& Yes & 54.36M & \cellcolor{lightgray}90.92\% & \cellcolor{lightgray}89.10\% & \cellcolor{lightgray}12.21\%\\
\midrule
\multirow{2}{*}{MobileNetV2} & No & 2.266M & \cellcolor{lightgray}89.88\% & \cellcolor{lightgray}87.86\% & \cellcolor{lightgray}18.31\%\\\cmidrule{2-6}
& Yes & 2.276M & 84.18\% & 81.04\% & 33.33\%\\
\midrule
\multirow{2}{*}{VIB\_SD} & No & 1.745M & 92.85\%  & 91.42\% & 15.50\% \\\cmidrule{2-6}
& Yes & 1.897M & \cellcolor{lightgray}94.14\% & \cellcolor{lightgray}92.96\% & \cellcolor{lightgray}13.15\%\\
\bottomrule
\end{tabular}
\end{table*}

\begin{table*}
\renewcommand{\arraystretch}{1.1}
\setlength\tabcolsep{4pt}
\caption{Results of using the Landsat\_smk dataset}
\label{results_Landsat}
\centering
\begin{tabular}{c c c c c c}
\toprule
\textbf{Model} & \textbf{InAmp} & \textbf{\#Params} & \textbf{Accuracy} & \textbf{Kappa} & \textbf{FN}\\
\midrule
\multirow{2}{*}{ResNet50} & No & 23.60M & 75.82\% & 63.68\% & \cellcolor{lightgray}26.47\%\\\cmidrule{2-6}
& Yes & 23.69M & \cellcolor{lightgray}80.43\% & \cellcolor{lightgray}70.70\% & 29.41\%\\
\midrule
\multirow{2}{*}{InceptionResnetV2} & No & 54.34M & 83.97\% & 75.89\% & 24.77\%\\\cmidrule{2-6}
& Yes & 54.35M & \cellcolor{lightgray}85.05\% & \cellcolor{lightgray}77.52\% & \cellcolor{lightgray}18.38\%\\
\midrule
\multirow{2}{*}{MobileNetV2} & No & 2.263M & 76.90\% &  65.11\% &  22.06\% \\\cmidrule{2-6}
& Yes & 2.272M & \cellcolor{lightgray}78.80\% & \cellcolor{lightgray}68.01\% & \cellcolor{lightgray}21.32\%\\
\midrule
\multirow{2}{*}{VIB\_SD} & No & 1.676M & 81.79\% & 72.61\% & 24.26\%\\\cmidrule{2-6}
& Yes & 1.812M & \cellcolor{lightgray}85.33\% & \cellcolor{lightgray}77.87\% & \cellcolor{lightgray}13.97\%\\
\bottomrule
\end{tabular}
\end{table*}

\subsection{Visualisation of InAmp-extracted Spectral Patterns}\label{InAmp_patterns}

To better understand the spectral patterns extracted by the InAmp module, we visualised some of the class-specific spectral patterns extracted by the VIB\_SD model from imagery samples in both the USTC\_SmokeRS and Landsat\_smk datasets. 

\begin{figure}[!h]
\centering
\includegraphics[width=2.5in]{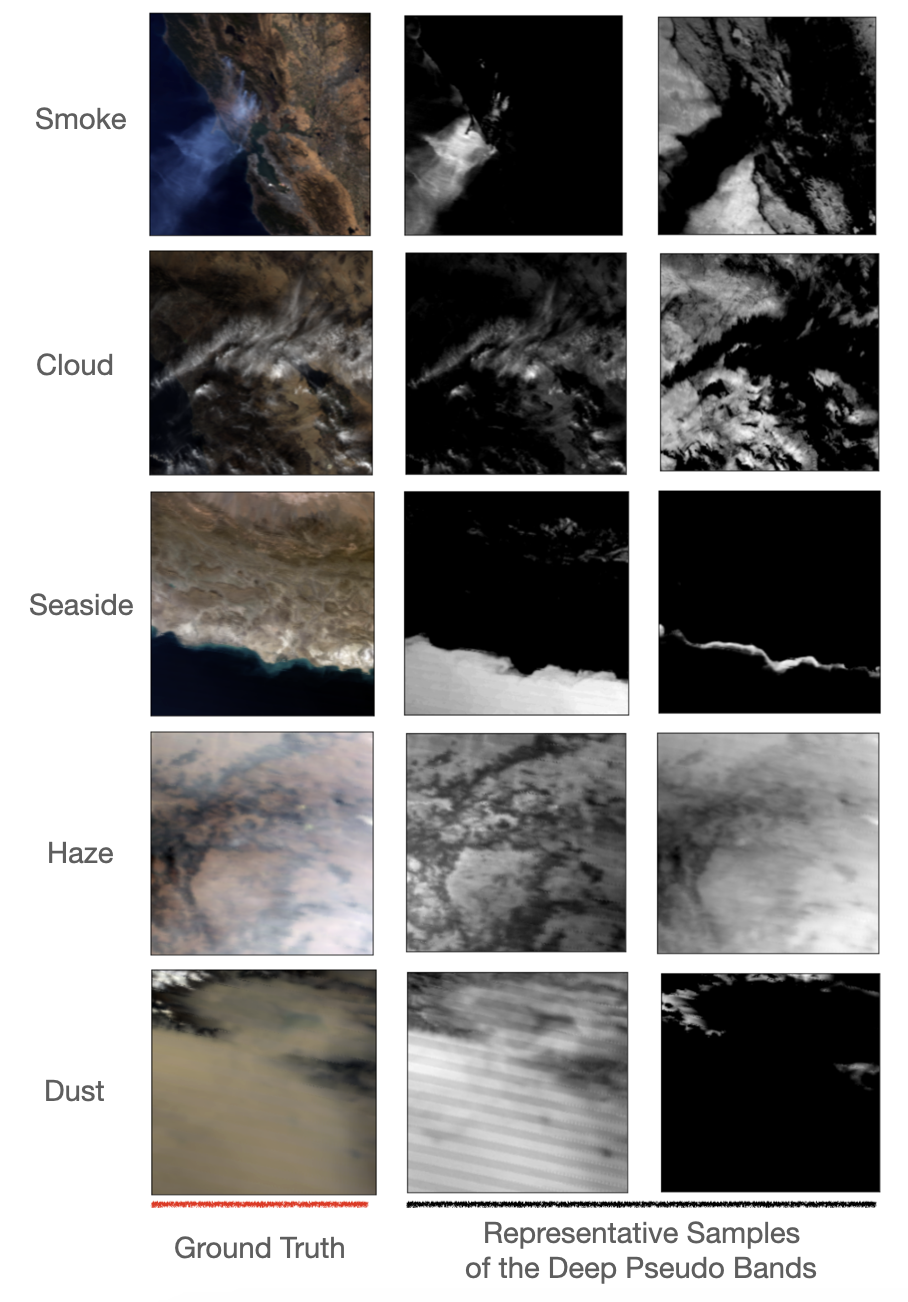}
\caption{Original MODIS imagery samples (left) vs. two samples of deep-pseudo bands extracted by InAmp}
\label{Modis_pseudo_bands}
\end{figure}

In Figure~\ref{Modis_pseudo_bands}, the five imagery samples in the leftmost column are ``Cloud'', ``Dust'', ``Haze'', ``Seaside'', and ``Smoke'' from the USTC\_SmokeRS dataset; the grey-scale images in the two columns on the right are the visualisation of two corresponding deep-pseudo bands from the output of the InAmp module. We observe that the spectral patterns present class-specific attributes. The pixels either representing or belonging to the target class, or belonging to other classes, were highlighted respectively in the deep-pseudo bands. Particularly, Figure~\ref{Modis_pseudo_bands} demonstrated that the InAmp module can precisely mark the smoke pixels even without the pixel-level ground truth.

\begin{figure}[!h]
\centering
\includegraphics[width=2.5in]{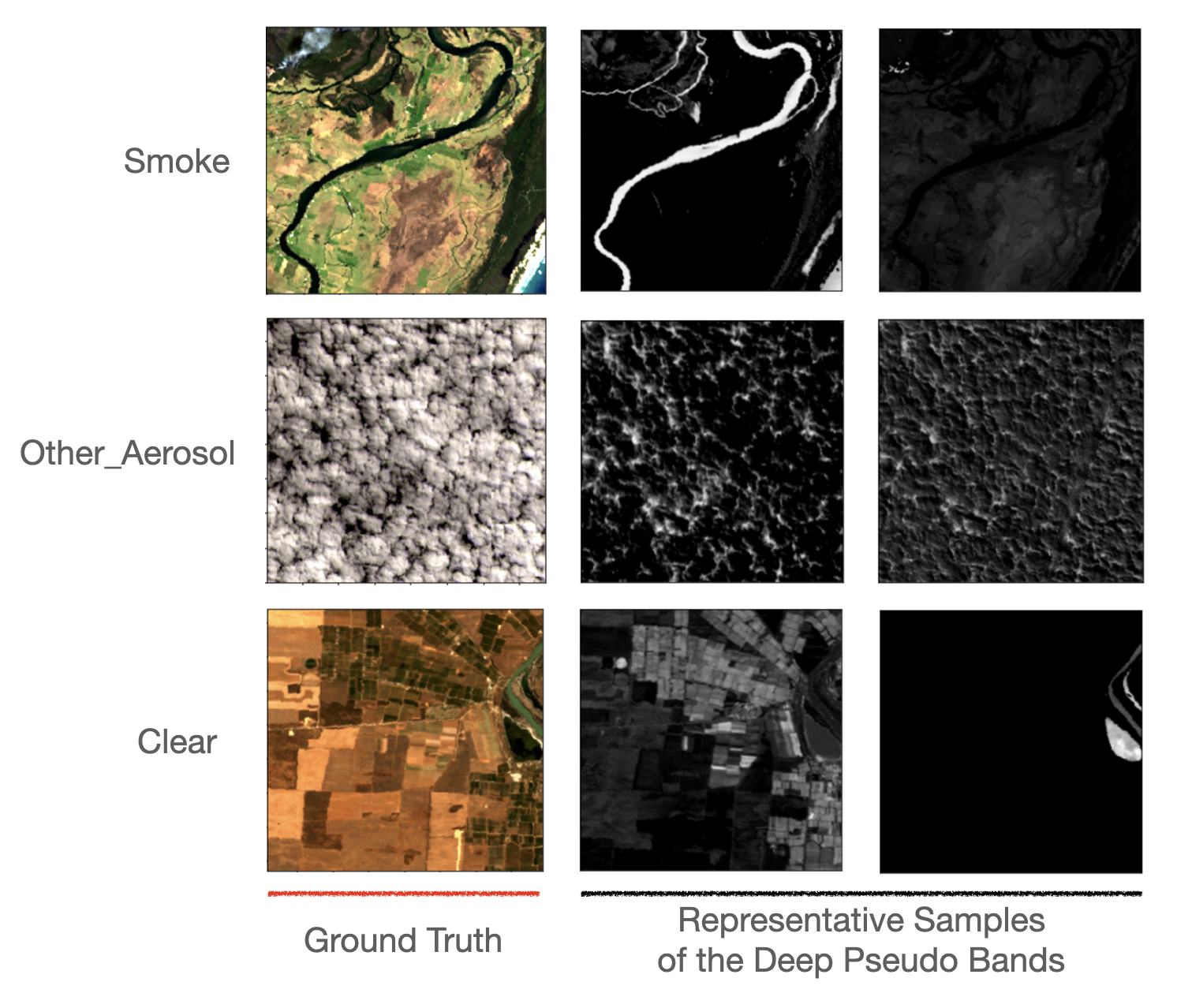}
\caption{Original Landsat imagery samples (left) vs. two samples of deep-pseudo bands extracted by InAmp}
\label{Ls_pseudo_bands}
\end{figure}

In Figure~\ref{Ls_pseudo_bands}, on the left are three ground truth imagery samples labelled as ``Smoke'', ``Other\_aerosol'', and ``Clear'' from the Landsat\_smk dataset, visualised with the RGB bands; on the right are the visualisation of two corresponding spectral patterns containing class-specific information. It can be observed that the spectral patterns capture class specific pixels very well for both fire smoke and other aerosols.

Figure~\ref{Modis_pseudo_bands} and Figure~\ref{Ls_pseudo_bands} also demonstrate that the InAmp module can make the model more explainable. The visualisation of the deep-pseudo bands showed evidences of what spectral patterns were learned and adopted by the model to make the prediction.

\subsection{Ablation Studies and Parameter Selection}\label{ablation}

We conducted ablation studies to determine how to integrate the attention mechanism in the InAmp module, how many $1 \times 1$ Conv2D layers the InAmp module should have, and how many channels the InAmp module should output.

We only used the VIB\_SD model for the ablation studies. In terms of the ablation study about the attention mechanism and the number of $1 \times 1$ convolution layers, we only used the USTC\_SmokeRS dataset. We used both datasets in terms of the ablation study about the number of output channels of the InAmp module.

The results about the attention mechanism are shown in Table~\ref{ablation_InAmp}, which suggests that the VIB\_SD employing the InAmp module with both spatial and channel attention achieved the highest accuracy.

\begin{table}%[width=.9\linewidth,cols=4,pos=h]
\renewcommand{\arraystretch}{1.2}
\caption{Ablation study about attention mechanism}
\label{ablation_InAmp}
\centering
\begin{tabular}{c c c c}
\toprule
\textbf{Attention} & \textbf{Accuracy} & \textbf{Kappa} & \textbf{FN}\\
\midrule
None &  92.85\% & 91.42\% & 15.50\%\\
CA &  91.57\% & 89.88\% & 16.43\%\\
SA &  92.29\% & 90.74\% & 18.31\%\\
CA \& SA & \cellcolor{lightgray}94.14\% & \cellcolor{lightgray}92.96\% & \cellcolor{lightgray}13.15\%\\
\bottomrule
\end{tabular}
\end{table}

The results regarding the $1 \times 1$ Conv2D layers are shown in Table~\ref{ablation_ConvNum}, which suggests that using two $1 \times 1$ Conv2D layers achieved the highest accuracy.

\begin{table}%[width=.9\linewidth,cols=4,pos=h]
\renewcommand{\arraystretch}{1.2}
\caption{Ablation study about the number of $1 \times 1$ Conv2D Layers}
\label{ablation_ConvNum}
\centering
\begin{tabular}{c c c c}
\toprule
\textbf{\#$1\times1$ Conv2D Layers} & \textbf{Accuracy} & \textbf{Kappa} & \textbf{FN}\\
\midrule
1 &  92.37\% & 90.84\% & 16.43\%\\
2 &  \cellcolor{lightgray}94.14\% & \cellcolor{lightgray}92.96\% & 13.15\%\\
3 &  93.98\% & 92.77\% & \cellcolor{lightgray}11.27\%\\
4 &  91.89\% & 90.26\% & 18.31\%\\
\bottomrule
\end{tabular}
\end{table}

The results about the number of output channels of the InAmp module are shown in Table~\ref{ablation_ChannelNum}. Considering the ablation study results obtained from both datasets and the computing complexity, we adopted 32 output channels in the proposed InAmp module.

\begin{table*}
\renewcommand{\arraystretch}{1.2}
\caption{Ablation study about the output channel number of the InAmp module}
\label{ablation_ChannelNum}
\centering
\begin{tabular}{c c c c c}
\toprule
\textbf{Dataset} & \textbf{\#Output Channels} & \textbf{Accuracy} & \textbf{Kappa} & \textbf{FN}\\
\midrule
\multirow{5}{*}{USTC\_SmokeRS}
& 16 &  92.37\% & 90.84\% & 11.74\%\\\cmidrule{2-5}
& 24 &  93.33\% & 92.00\% & 15.02\%\\\cmidrule{2-5}
& 32 &  \cellcolor{lightgray}94.14\% & \cellcolor{lightgray}92.96\% & 13.15\%\\\cmidrule{2-5}
& 40 &  \cellcolor{lightgray}94.14\% & \cellcolor{lightgray}92.96\% & \cellcolor{lightgray}10.33\%\\\cmidrule{2-5}
& 48 &  93.33\% & 92.00\% & 12.21\%\\
\midrule
\multirow{5}{*}{Landsat\_smk}
& 16 &  80.43\% & 70.53\% & 22.06\%\\\cmidrule{2-5}
& 24 &  82.88\% & 74.21\% & 16.91\%\\\cmidrule{2-5}
& 32 &  \cellcolor{lightgray}85.33\% & \cellcolor{lightgray}77.87\% & \cellcolor{lightgray}13.97\%\\\cmidrule{2-5}
& 40 &  82.07\% & 73.08\% & 18.38\%\\\cmidrule{2-5}
& 48 &  80.43\% & 70.41\% & 16.91\%\\
\bottomrule
\end{tabular}
\end{table*}

\section{Discussion} \label{discussion}

Fire smoke shares some similar spatial patterns with clouds, haze, fog, etc. This makes it challenging for discerning smoke from these aerosols with DL models that mainly rely on spatial patterns. However, the particles in different aerosols have distinct reflection characteristics to different bands, which makes spectral patterns useful in fire smoke detection. We believe this is the major reason why the InAmp module improves the accuracy of the CNN models for fire smoke detection in satellite imagery.  

We observed that the performance of integrating the InAmp module with different CNN architectures may be influenced by the specific architectures. For instance, when we trained MobileNetV2 with the InAmp module using the USTC\_SmokeRS dataset, we did not observe any improvement in performance. We hypothesise that this could be due to the extensive use of depth-wise separable convolution and inverted residual blocks in MobileNetV2, and both the depth-wise separable convolution and inverted residual blocks contain $1 \times 1$ Conv2D layers. As a result, these $1 \times 1$ Conv2D layers could potentially distort some of the spectral patterns learned by the InAmp module, which in turn generate cross-channel associations using $1 \times 1$ Conv2D layers. This distortion could lead to a compromised performance. For a more detailed explanation of the MobileNetV2 architecture, readers can refer to \cite{sandler2018mobilenetv2}. Further investigations are required to identify the root cause of the reduced performance and explore potential solutions.

Since the InAmp module serves as an input pre-processing block, it is easily applicable to ViTs apart from CNNs. Nonetheless, we were unable to evaluate its efficacy on ViTs in this study due to the difficulty in identifying, implementing, and training appropriate benchmark ViTs within the constraints of our research project. We plan to conduct additional research to explore this avenue in the future.  

The InAmp module is not limited to fire smoke detection from satellite imagery and can be applied to a wider range of classification tasks, including those using non-satellite imagery. In particular, it would be useful for tasks where the reflection characteristics of a class in different bands are distinct from those of other classes. For example, detecting water pollution, vegetation diseases, or diagnosing human diseases such as polyps or skin cancer may benefit from the InAmp module's ability to extract class-specific spectral patterns. Future work could investigate the effectiveness of the InAmp module on such tasks, as it is beyond the scope of our current research.

In addition, we noticed that some of the deep-pseudo bands extracted by the InAmp module can successfully mark the pixels belonging to certain target classes, even no binary ground truths at the pixel level were provided. This suggests the potential of using the InAmp module for tasks involving pixel-level labelling or segmentation. Besides, since the deep-pseudo bands extracted by the InAmp module can be easily visualised, they can enhance the interpretability of the DL models.

Furthermore, the InAmp module can be leveraged to facilitate transfer learning for training a fire smoke detection CNN model using imagery from multiple satellite sensors or updating a trained model with a few labelled images from a new satellite sensor. For instance, one can first train a CNN model on a dataset with fewer spectral bands and then add the InAmp module in front of the trained model, setting the input channels to the number of bands in the new dataset and output channels to the number of bands in the original dataset. This will allow the new model to be fine-tuned through transfer learning using only a small number of images in the new dataset. 

\section{Conclusion} \label{conclusion}

In conclusion, our study demonstrates the limitations of current DL models in exploring pixel-level spectral patterns that are critical for fire smoke detection in satellite imagery. To address this, we have proposed a novel DL module called InAmp that enables DL models to extract class-specific pixel-level spectral patterns alongside spatial patterns. The InAmp module incorporates $1 \times 1$ filters and attention mechanisms and can be seamlessly integrated with existing DL models with minimal computational overhead.

We evaluated the InAmp module on two fire smoke satellite imagery datasets: USTC\_SmokeRS with only the RGB bands and Landsat\_smk with six spectral bands (i.e., RGB, NIR, SWIR\_1, and SWIR\_2). The experimental results demonstrate that integrating the InAmp module with an existing CNN model effectively improves the model's prediction accuracy.

We also visualised the deep-pseudo bands extracted by the InAmp module, and demonstrated that the deep-pseudo bands successfully segmented pixels belonging to specific target classes. This suggests the potential of using the InAmp module for pixel-level labeling or segmentation tasks.

Moreover, the InAmp module learns to extract class-specific pixel-level spectral patterns during the learning process based on the classification tasks, indicating its potential to be applied to a broader range of classification tasks in various domains. Overall, the InAmp module is a promising approach to improving DL models' interpretability and accuracy, particularly in multi-spectral satellite imagery analysis.

\section*{Acknowledgment}

This work has been supported under project P3-07s by the SmartSat CRC, whose activities are funded by the Australian Government’s CRC Program.

\bibliographystyle{unsrt}  
\bibliography{references}

\end{document}